\documentclass[12pt]{article}

\usepackage{eqnarray,amsmath}
\usepackage{amssymb}
\usepackage{cancel}
\usepackage{algorithm}
\usepackage{algpseudocode}

\usepackage{fourier}
\usepackage[normalem]{ulem}
\usepackage[T1]{fontenc}

\usepackage{accents}
\usepackage{sectsty}
\sectionfont{\fontsize{12}{15}\selectfont}
\subsectionfont{\normalfont\fontsize{12}{15}\selectfont}
\subsubsectionfont{\itshape\normalfont\fontsize{12}{15}\selectfont}
\usepackage{bm}
\usepackage{lineno}

\usepackage{relsize}

\DeclareMathOperator{\expit}{expit}

\usepackage{accents}
\newlength{\dhatheight}

\newcommand*\patchAmsMathEnvironmentForLineno[1]{%
	\expandafter\let\csname old#1\expandafter\endcsname\csname #1\endcsname
	\expandafter\let\csname oldend#1\expandafter\endcsname\csname end#1\endcsname
	\renewenvironment{#1}%
	{\linenomath\csname old#1\endcsname}%
	{\csname oldend#1\endcsname\endlinenomath}%
}
\newcommand*\patchBothAmsMathEnvironmentsForLineno[1]{%
	\patchAmsMathEnvironmentForLineno{#1}%
	\patchAmsMathEnvironmentForLineno{#1*}%
}
\AtBeginDocument{%
	\patchBothAmsMathEnvironmentsForLineno{equation}%
	\patchBothAmsMathEnvironmentsForLineno{align}%
	\patchBothAmsMathEnvironmentsForLineno{flalign}%
	\patchBothAmsMathEnvironmentsForLineno{alignat}%
	\patchBothAmsMathEnvironmentsForLineno{gather}%
	\patchBothAmsMathEnvironmentsForLineno{multline}%
}

\makeatletter
\newenvironment{megaalgorithm}[1][htb]{%
    \renewcommand{\ALG@name}{Box}
   \begin{algorithm}[#1]%
  }{\end{algorithm}}
\makeatother

\usepackage{color}
\usepackage{wrapfig}
\usepackage{indentfirst} 
\usepackage{setspace}
\usepackage[numbers,super,sort&compress]{natbib} 
\usepackage{booktabs} 
\usepackage{rotating} 
\usepackage[margin = .85in]{geometry} 
\usepackage{fancyhdr} 
\usepackage{tabularx}
\usepackage{pdfpages}
\usepackage{longtable}

\usepackage{subfig}
\usepackage{mdwlist}
\usepackage{url}
\usepackage{verbatim}
\usepackage{multirow}
\usepackage{multicol}
\usepackage[compact]{titlesec}
\usepackage[colorlinks = TRUE, urlcolor = blue, linkcolor = black, citecolor = black]{hyperref}
\usepackage{float}
\usepackage{lscape}
\usepackage{graphicx}
\makeatletter
\renewcommand\@biblabel[1]{#1.}
\makeatother

\graphicspath{{../figures/}}

\begin{document}
\thispagestyle{empty}
\baselineskip=28pt

\noindent Brief Report

\vskip .2cm
{{\noindent \huge Computing True Parameter Values in Simulation Studies Using Monte Carlo Integration}}

\baselineskip=12pt

\vskip .5cm
\noindent Ashley I. Naimi, PhD$^{1, *}$\\[.5em]
\noindent David Benkeser, PhD$^{2}$\\[.5em]
\noindent Jacqueline E. Rudolph, PhD$^{3}$\\[.5em]

\vskip .5cm
\noindent $^1$ Department of Epidemiology, Emory University.\\[.5em]
\noindent $^2$ Department of Biostatistics, Emory University.\\[.5em]
\noindent $^3$ Department of Epidemiology, Johns Hopkins University.\\[.5em]

\vskip .5cm
\noindent \hskip -.2cm
\begin{tabular}{ll}
*Correspondence: & Department of Epidemiology \\[-.1cm]
& Rollins School of Public Health \\[-.1cm]
& Emory University \\[-.1cm]
& 1518 Clifton Road \\[-.1cm]
& Atlanta, GA 30322\\[-.1cm]
& \href{mailto:ashley.naimi@emory.edu}{ashley.naimi@emory.edu}
\end{tabular}
\vfill 
\vskip .5cm
\noindent Conflicts of Interest: None
\vskip .5cm
\noindent Acknowledgements: We thank Dr Tim Morris for helpful comments on a previous version of this manuscript.
\vskip .5cm
\noindent Sources of Funding: AIN was supported by NIH grant number R01	HD102313.
\vskip .5cm

\noindent \hskip -.2cm
\begin{tabular}{ll}
Target Journal: 	& \emph{Epidemiol} \\[-.01cm]
Text word count: 	& 1,800 \\[-.01cm]
Abstract word count: &  178 \\[-.01cm]
Number of Figures: 	& 2 \\[-.01cm]
Number of Tables:  &	0  \\[-.01cm]
Number of References:  & 15	  \\[-.01cm]
Running head:  &	 Monte Carlo Integration for Simulation Estimands \\ 
\end{tabular}
%
%
\newpage
\thispagestyle{empty}
\begin{center}
{\Large{\bf Abstract}}
\end{center}
\baselineskip=12pt

\noindent  
Simulation studies are used to evaluate and compare the properties of statistical 
methods in controlled experimental settings. In most cases, performing a simulation study requires knowledge of the 
true value of the parameter, or estimand, of interest. However, in many simulation designs, the true value of the estimand is difficult to compute analytically. Here, we illustrate the use of Monte Carlo integration to compute true estimand 
values in simple and more complex simulation designs. We provide general pseudocode that can be replicated in any 
software program of choice to demonstrate key principles in using Monte Carlo integration in two scenarios: a simple 
three variable simulation where interest lies in the marginally adjusted odds ratio; and a more complex causal mediation 
analysis where interest lies in the controlled direct effect in the presence of mediator-outcome confounders affected 
by the exposure. We discuss general strategies that can be used to minimize Monte Carlo error, and to serve as checks 
on the simulation program to avoid coding errors. R programming code is provided illustrating the application of our 
pseudocode in these settings.
%
%
%
\baselineskip=12pt
\par\vfill\noindent

\par\medskip\noindent
\newpage
\doublespacing
\setcounter{page}{1}

\section*{Introduction}

Simulation studies are often employed to evaluate and compare the properties of statistical 
methods in controlled experimental settings. Simulation studies can be used for basic conceptual clarification,\cite{Rudolph2021a} to formal research questions about the performance of different analytic methods.\cite{Morris2019} Monte Carlo simulations relate to a general category of techniques known as Monte Carlo methods,\cite{Metropolis1949} a class of methods that rely on pseudo-random number generation to solve problems.\cite{McCracken1955}

The design of Monte Carlo simulation studies involves several steps. These include clearly articulating the aims of the simulation study, the data generating mechanisms that will be used, the estimand of interest, the methods that will be used to estimate the estimand, and the performance measures used to evaluate the properties of the methods under study.\cite{Morris2019} Simulation studies proceed by repeatedly simulating data sets from a data generating mechanism. Each simulated data set is analyzed, and the performance the methods is compared by aggregating results across multiple simulated data sets. For example, to approximate the bias of an estimator, we can compute the empirical average (across all simulated data sets) difference between the point estimates and the truth. To approximate the coverage of a method for building a confidence interval, we can compute the empirical proportion (out of all simulated data sets) of intervals that contain the true parameter value. This makes clear the importance for the researcher to know the true value of the estimand that is implied by the selected data generating process. 

Ideally, the true estimand value will be taken directly from the parameter values used in the 
data generating mechanism. For example, if the data generating mechanism involves simulating from a 
logistic regression $P(Y = 1 \mid A, C) = \mbox{expit}\{\beta_0 + \beta_1 A + \beta_2 C\}$, with $\mbox{expit}\{\bullet\} = \frac{1}{(1+\exp(-\bullet))}$, and interest lies in estimating the conditionally adjusted odds ratio for $A$, the true estimand value is easily computed as $\mbox{exp}(\beta_1)$, where $\beta_1$ is the researcher-selected value used to generate the data. 

It is sometimes possible to read off true estimand values directly from the data generating process. However, often the true estimand value may not be easily computed as a single parameter, or simple combination of parameters, used to generate the data. Here, we outline the use of Monte Carlo integration, a process that utilizes pseudo-random sampling to approximate complex analytic expressions, to solve for true estimand values. We illustrate the approach using two example scenarios: a data generating mechanism where the estimand of interest is the marginally adjusted odds ratio; and a causal mediation setting where interest lies in the controlled direct effect. We provide pseudocode for each example, and provide R code in an associated \href{https://github.com/ainaimi/MonteCarloIntegration}{GitHub} repository.

\section*{Example 1: Marginally Adjusted Odds Ratio as the Estimand}

The first example involves estimating the marginally adjusted odds ratio using methods such as inverse probability weighting or marginal standardization \cite{Naimi2017, Naimi2020}. The causal diagram is a simple triangle structure with $C$ as the confounder, $A$ as the exposure, and $Y$ as the outcome. A parametric approach to simulating such data might be to first simulate $C$ from a Normal distribution with mean $\mu$ and variance $\sigma^2$. Then based on the simulated values of $C$, simulate the exposure $A$ from a logistic regression model with $P(A = 1 \mid C) = \mbox{expit}(\alpha_0 + \alpha_1 C)$, for some researcher-selected values $\alpha_0, \alpha_1$. Finally, the outcome is simulated from a logistic regression model \begin{equation}
P(Y = 1 \mid X, C) = \expit \left \{ \beta_0 + \beta_1 A + \beta_2 C \right \} \ .
\label{eq:logisticModel}
\end{equation}

As noted above, if the true parameter of interest was the conditionally adjusted odds ratio, its value could easily be obtained as $\exp (\beta_1)$. However, are well known and important differences between marginally versus conditionally adjusted odds 
ratios,\cite{Neuhaus1993,Daniel2021,Pang2013} due to the fact that the odds ratio is a noncollapsible measure of effect.\cite{Greenland2005b} Noncollapsibility is distinct from confounding,\cite{Pang2013a} which means that for a given logistic regression model with the same set of correctly chosen confounders, the marginally versus conditionally
adjusted odds ratio may be numerically different. Thus, in the context of equation \ref{eq:logisticModel}, 
while exponentiating the value of the $\beta_1$ coefficient provides the conditionally
adjusted odds ratio, the marginally adjusted odds ratio may be a different value. 

Under ideal conditions, one would rely on algebraic derivations to obtain an analytic solution to the marginally adjusted 
odds ratio. Doing this for equation \ref{eq:logisticModel} could be accomplished via integration, which would require solving for:

$$\mu(a) = \int_c \expit \{ \beta_0 + \beta_1 a + \beta_2 c \} \frac{1}{(2\pi)^{1/2} \sigma} \mbox{exp}\left\{ \frac{(c - \mu)^2}{2\sigma^2} \right\} dc \ ,$$

for $a = 0, 1$. The solution to this integral yields a marginal probability of the outcome that would be observed if $A = a$. One can then construct a marginally adjusted odds ratio as:

$$\psi = \frac{\mu(1)}{1 - \mu(1)} \bigg/ \frac{\mu(0)}{1 - \mu(0)}$$

However, even in this simple setting with a single variable $C$, analytically solving for this integral is challenging. One could employ numeric integration (See \href{https://github.com/ainaimi/MonteCarloIntegration}{GitHub} repository). However, this approach would not scale well to settings with more than one $C$ variable. 

Instead, we can use Monte Carlo integration to solve for $\psi$, as follows:

\begin{megaalgorithm}
\caption{Pseudocode for Implementing Monte Carlo Integration to Solve for the True Marginally Adjusted Odds Ratio in a Simple Three Node Causal Diagram}\label{alg:noncollapsibility_algorithm}
\begin{algorithmic}[1]

	\State set random number generator seed value
	
	\State set large sample size $N$

	\State simulate $i \in 1 \ldots N$ observations $C_i$ from a Normal distribution with mean $\mu$ and variance $\sigma^2$

	\State compute $\hat{\mu}_i(a) = \mbox{expit}(\beta_0 + \beta_1 a + \beta_2 C_i)$ for both $a = 0$ and $a = 1$. Thus, $\hat{\mu}_i(0) = \expit \left ( \beta_0 + \beta_2 C_i \right )$ and $\hat{\mu}_i(1) = \expit \left ( \beta_0 + \beta_1 + \beta_2 C_i \right )$.

	\State the approximated value of $\mu(x)$ is given by the mean of $\hat{\mu}_i(x)$ over all $N$ simulated observations, $\hat{\mu}(x) = \frac{1}{N} \sum_{i = 1}^N \hat{\mu}_i(x)$. The approximated value of $\psi$ is given by
		$$\hat{\psi} = \frac{ \left [ \hat{\mu}(1)/(1 - \hat{\mu}(1)) \right ]}{\left [ \hat{\mu}(0)/(1 - \hat{\mu}(0)) \right ]}$$

\end{algorithmic}
\end{megaalgorithm}

The approximated value of $\psi$ obtained from the pseudocode above can be used as the true estimand value for a simulation study evaluating the properties of an estimator seeking to quantify the marginally adjusted odds ratio. However, it is important to note that this "true" estimand value depends on: (i) the specifications for the distribution of $C$; and (ii) the parameter values for the logistic regression model generating $Y$ in this case, $\beta_0, \beta_1, \beta_2$. If the distribution of $C$ changes, or if the regression model coefficients change, this pseudoalgorithm should be run again under these new settings. Additionally, the true value of $\psi$ will be subject to error that depends on the sample size used in the Monte Carlo integration. This can have important consequences for simulation results. To mitigate the impact of Monte Carlo error, a number of strategies could be entertained. First, the sample size $N$ used to compute the true value should be as large as possible under the computing constraints to minimize Monte Carlo error in the true value. Second, one could more comprehensively assess the variance of the Monte Carlo error of the true value under different $N$ to ensure that it is acceptably small in the region of $N$ being used to compute the truth. Lastly, several iterations of the pseudo-code under different seed values could be run to confirm that the sample is large enough to lead to only negligible changes in the approximated value.

Of note, in step 5 of the above pseudocode, one should compute $\hat{\mu}_i(a = 0)$ and $\hat{\mu}_i(a=1)$
directly from the logistic model for the outcome, and not to convert these probabilities to binary outcomes with a draw from the Bernoulli distribution. Doing so prevents the introduction of additional Monte Carlo error, and reduces the variability of the Monte Carlo integration method. 

\section*{Example 2: Controlled Direct Effect as the Estimand}

The second example computes the true controlled direct effect (CDE) value. Here, the CDE of interest on the difference scale:

$$\psi = E(Y^{a, m} - Y^{a^{\star}, m})$$

\noindent where $Y^{a, m}$ is the potential outcome that would be observed if the exposure $A$ were set to some value $a$ and the mediator $M$ were set to some value $m$, while $Y^{a^{\star}, m})$ is the corresponding value that would be observed under some referent $a^{\star}$. 

\begin{figure}[H]
\par\noindent\rule{\textwidth}{0.4pt}
\caption{Causal diagram depicting the relationships between an exposure $A$, a mediator $M$, a confounder $C$, a 
mediator-outcome confounder affected by the exposure $L$, and an outcome $Y$.}
\begin{center} 
	\includegraphics[scale=1]{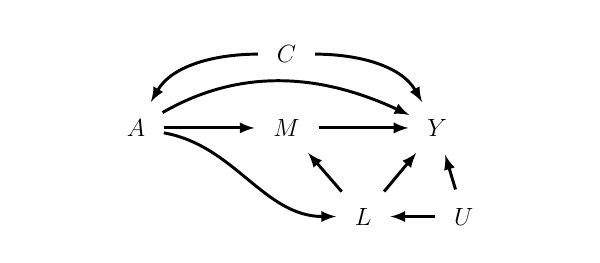}
\end{center}
\par\noindent\rule{\textwidth}{0.4pt}
\label{fig:mediationdag}
\end{figure}

A causal diagram representing our data generating mechanism is displayed in Figure \ref{fig:mediationdag}.

In this example, the true value of the controlled direct effect is difficult to compute analytically. It consists of a 
a function of the magnitudes of the following paths in Figure \ref{fig:mediationdag}:

$$A \rightarrow Y$$
$$A \rightarrow M \rightarrow Y$$
$$A \rightarrow L \rightarrow Y$$

In many such settings, it simply may not be possible to analytically solve for the true controlled direct effect. However, 
we can use Monte Carlo integration as follows:

\begin{megaalgorithm}
	\caption{Pseudocode for Implementing Monte Carlo Integration to Solve for the True Controlled Direct Effect in a Causal Mediation Analysis DAG with a Mediator-Outcome Confounder Affected by the Exposure.}\label{alg:mediation_algorithm}
	\begin{algorithmic}[1]
	
		\State set random number generator seed value.
		
		\State set large sample size $N$.
	
		\State simulate $i \in 1 \ldots N$ observations $C_i$ and $U_i$ from a distribution of choice.
	
		\State construct $a_1$ and $a_0$ for all $N$ observations corresponding to exposed (e.g., $a_1 = 1$) and referent (e.g., $a_0 = 0$) states.

		\State simulate two $L_i$ variables, one corresponding to $a_1$ ($L^{a_1}_i$) and one corresponding to $a_0$ ($L^{a_0}_i$), for all $N$ observations. This 
		would be from a regression model that includes the simulated $U_i$ and the specified $a_x$ value, along with relevant parameters.

		\State construct $m$ for all $N$ observations corresponding to the specific mediator value of interest in the controlled direct effect contrast (e.g., $m = 0$ for all $N$).
	
		\State simulate $Q^{a_x, m}_i$ for all $i \in 1 \ldots N$ observations from the model for the outcome:
			$$Q^{a_x, m}_i = f \left ( \beta_0 + \beta_1 a_x + \beta_2 C_i + \beta_3 m + \beta_4 L_i + \beta_5 U_i \right ),$$

			where $f(\bullet)$ represents a specified link function of interest.
	
		\State take the mean of $Q^{a_x, m}_i$ over all $N$ to obtain $\mu_{a_x, m} = \frac{1}{N} \sum_{i = 1}^N Q^{a_x, m}_i$, and construct a difference measure as:
			$$\widetilde{\psi} = \mu_{a_1, m} - \mu_{a_0, m}$$
	
	\end{algorithmic}
\end{megaalgorithm}

Again, the value of $\widetilde{\psi}$ obtained from the psuedocode above can be used as the true estimand value for the controlled direct effect under the specified simulation settings. The magnitude of this true value will again depend on the specific parameter values, distributions, and functional forms selected to generate all variables used in the outcome model. If any of these are  changed, a separate  $\widetilde{\psi}$ should be quantified. 

\section*{Discussion}

Monte Carlo simulation studies are commonly used to evaluate estimators in a range of scientific 
settings. However, estimator properties such as bias, mean squared error, and confidence interval coverage require a numerical value for the true estimand of interest. In many settings, this true value cannot be obtained analytically. Several simulation studies using Monte Carlo integration to compute the true parameter values have been done.\cite{Franklin2015, Naimi2014} However, to our knowledge, there is no practical description of the procedure, or the setbacks that should be considered when deploying the approach, that can serve as a reference for researchers unfamiliar with the technique. 

There are several important issues to consider when designing a program to implement Monte Carlo integration to compute 
estimand values. First, general recommendations suggest that the selected sample size used to compute the true 
value should be as large as is computationally feasible to minimize Monte Carlo error to the smallest value possible. 
However, in practice a degree of Monte Carlo error could be tolerated. For example, if increasing the Monte Carlo sample
size past a certain boundary $\kappa$ only changes the estimand value at the 5th or greater decimal place, it might be advisable to set the Monte Carlo sample size to $\kappa$ to avoid expending resources for little meaningful gain. Second, it is important to note that the source of Monte Carlo error in any simulation study is derived from programming calls to the 
system's random number generator. For example, in the R programming language, functions used to generate observations 
from a distribution of interest [e.g., rnorm(), runif(), or rexp()], or functions used to sample from a set of observations, 
will contribute to Monte Carlo error. For this reason, one strategy that should be used to minimize Monte Carlo error in the
estimand value is to minimize these calls where possible (e.g., Step 5 in Algorithm 1, or Step 7 in Algorithm 2). 

Several classes of simulation study design exist. In fully parametric simulation studies, each component of the data 
generating mechaism is specified parametrically using researcher defined functions. For example, confounders, exposures, and 
outcomes will all be drawn from known distributions with pre-specified relationships between each. Alternatively, plasmode
simulations can be conducted in which a portion of the data generating mechanism (e.g., confounders) can be constructed using actual data sampled from a population of interest.\cite{Franklin2014} Doing so preserves the empirical associations between variables drawn from the sample, thus providing a closer approximation between simulated and real data. To compute the true value using Monte Carlo integration with a plasmode design using data from a fixed sample size, one could resample the original data with replacement to obtain a sufficiently large sample that will minimize Monte Carlo error. Finally, synthetic simulations have recently been developed that rely on machine learning to simulate data whose variables and joint distributions between them closely approximate an index dataset sampled from a real population of interest, but where the effect of the exposure of interest is specified \emph{a priori}.\cite{Athey2021,Parikh2022} In principle, Monte Carlo integration can be used in all three settings to derive the true estimand of interest, or to serve as a verification step in the process of simulating complex data. Indeed, the Monte Carlo method is a general methodological strategy that has been used across the sciences to solve simple and complex problems. With increasingly complex simulation analyses being carried out in epidemiology, Monte Carlo intergation will become increasingly useful for gaining insights into complex questions.

\end{document}